\input epsf
\documentstyle[preprint,aps,prb]{revtex}
\begin{document}

\draft

\title{
Non-magnetic impurities in two dimensional superconductors
}

\author{ T. Xiang and J. M. Wheatley}

\address{
Research Center in Superconductivity, University of Cambridge,
Madingley Road, Cambridge CB3 0HE, United Kingdom
}
\date{\today}

\maketitle
\begin{abstract}
A numerical approach to disordered 2D superconductors
described by BCS mean field theory is outlined.
The energy gap and the
superfluid density at zero temperature and the quasiparticle density
of states are studied. The method involves approximate self-consistent
solutions of the Bogolubov-de$\,$Gennes equations on finite
square lattices. Where comparison is possible, the
results of standard analytic approaches to this problem are reproduced.
Detailed modeling of impurity effects is practical
using this approach. The {\it range} of the impurity potential
is shown to be of {\it quantitative importance} in the case of
strong potential scatterers. We discuss
the implications for experiments, such as the rapid suppression of
superconductivity by Zn doping in Copper-Oxide superconductors.
\end{abstract}

\section{Introduction}

It is well established that chemical substitutions on copper oxide
superconductors can qualitatively alter the properties of
these materials in both their normal and superconducting
states\cite{Exp}.
In the normal state,impurity substitution can be
used to test models of electronic transport.
In the superconducting state, impurity scattering
effects are sensitive
to order parameter symmetry\cite{Theory} as well
as other properties.
It is well known that non-magnetic impurities are pair breakers
in d-wave or other non-trivial pairing states with nodes of
the energy gap on the Fermi surface. They produce a finite
lifetime of the quasiparticles around the gap nodes and
a finite density of states at low energy.

In order to parameterize impurity scattering,
a physical picture for the normal state, the superconducting state
and the impurity is required. In this paper we are motivated by Zn
doping experiments in cuprates
which we assume can be treated as a pure
potential scatterer of fermions which obey Luttinger's theorem.
Since d-orbitals of Zn$^{++}$ are fully
occupied, it is naively expected to behave as a non-magnetic impurity.
However, electronic correlations may modify this picture
somewhat. Nuclear Magnetic Resonance measurements
by Mahajan and co-workers\cite{NMR}
show that the copper spin correlations are
severely modified on neighboring Cu sites, which could in principle
give rise to effects analogous to the spin-flip scattering process of
conventional magnetic impurities.
As discussed recently by Borkowski and Hirschfeld\cite{BH}, the
unknown relative strength of spin flip and impurity scattering rates
in Zn doped systems poses an obstacle to the quantitative analysis of
experiments involving Zn even within the conventional BCS formalism.

Even if spin-flip scattering is negligible,
the modification of copper spin correlations
in the vicinity of an impurity site has the effect of
increasing the
range of an effective scattering
potential.
Information about the effective potential in
$YBa_2(Cu_{1-x}Zn_x)_3O_{6.9}$ can be obtained from the
residual sheet resistance estimated by
extrapolation from the normal state.
The residual resistance from two dimensional
potential scattering is
given by the approximate expression\cite{ziman}
\begin{equation}
\Delta\rho = {1\over \pi} x
{\displaystyle h\over \displaystyle e^2}
\sum_l\sin^2(\delta_l-\delta_{l+1}).
\label{rr}
\end{equation}
$\delta_l$ are the phase shifts which are constrained to satisfy the
Friedel sum-rule
$\Delta Z={2\over \pi}\sum_{l=-\infty}^\infty \delta_l$.
$\Delta Z$ is the difference in the number of conduction electrons
in the system without impurities and with one impurity.
If we assume Zn removes one electron from the conduction band
(i.e. $\Delta Z=-1$) and
the effective potential is an impenetrable disc of radius $a$,
we estimate that the residual sheet resistance
is $\approx 0.6k\Omega$ for $x=0.01$ in the $l=0$ dominant
scattering channel ($a\sim 0.54/k_f$),
which is a factor three smaller than the experimental
value $1.7k\Omega$\cite{RR}. This discrepancy may indicate
appreciable phase shifts in one or more higher
angular momentum channels.
In fact the $l=2$ ``near resonant channel", which requires a
larger radius of the scattering potential
($a\sim 2.8/k_f$)\cite{phaseshifts} yields
$\approx 1.8k\Omega$ for $x=0.01$. Recently Poilblanc,
Scalapino, and Hanke\cite{Poilblanc}
have investigated the effects of non-magnetic impurities in
antiferromagnetically correlated systems. They find that
the scatterings in $l\le 2$ channels are strong in that system.

The above discussion suggests that  detailed
structure of the impurity potential is important
in making a quantitative study of Zn doped materials.
As seen below, lattice effects may also be of
quantitative importance
for short coherence length superconductors.
Recently, in a brief
report\cite{Tao}, we presented a numerical study of
the disorder effect in two dimensional superconductors
of various pairing symmetries in
the limit of strong impurity potentials. We demonstrated that
a short but finite range potential has a much stronger effect than
an on-site (``$\delta$-function'') potential. In fact the
finite range of the potential may be the primary reason
for the rapid suppression of pairing correlations
with impurity concentration in $YBa(Cu,Zn)O$.
The importance of finite potential range has been pointed out
recently by Balatsky et al\cite{Balatsky} in connection with
non-universal behavior of the low frequency conductance in d-wave
superconductors.

In this paper we present a more detailed study of the disorder
effect in two dimensional superconductors.
The model used here is a lattice BCS mean-field Hamiltonian with
disorder defined by,
\begin{equation}
H[\Delta_{\mathbf r\tau} ]
=-t\sum_{\langle {\mathbf r}{\mathbf r}^\prime \rangle \sigma}
c_{{\mathbf r}\sigma}^\dagger c_{{\mathbf r}^\prime\sigma}
+\sum_{{\mathbf r}{\mathbf \tau}}
(\Delta_{{\mathbf r}{\mathbf \tau}}c_{{\mathbf r}\uparrow}^\dagger
c_{{\mathbf r}+{\mathbf \tau}\downarrow}^\dagger +
h.c.)+\sum_{{\mathbf r}\sigma}(\sum_{{\mathbf r}_{i}}
V_{{\mathbf r}_{i},\mathbf r}-\mu )
c_{{\mathbf r}\sigma}^\dagger c_{{\mathbf r}\sigma} ,
\label{ee1}
\end{equation}
where $\langle {\mathbf r}{\mathbf r}^\prime\rangle $
denotes nearest neighbors,
$\mu$ is the chemical potential. $V_{{\mathbf r}_{i},\mathbf r}$
is a scattering potential of an impurity at ${\mathbf r}_i$.
Lacking detailed knowledge of the scattering potential in high-$T_c$
cuprates, we assume a model potential form for it:
$V_{{\mathbf r}_i,\mathbf r}$=$V_0\delta_{{\mathbf r}_i,\mathbf r}
+V_1(\delta_{{\mathbf r}_i-{\mathbf r} \pm \hat x}
+\delta_{{\mathbf r}_i-{\mathbf r} \pm \hat y}) $.
When $V_1=0$, it is a $\delta$-function potential,
otherwise it is finite ranged.
$\Delta_{{\mathbf r}{\mathbf \tau}}$ is
the superconducting gap parameter.
For a given filling factor of electrons $n_e$,
$\Delta_{\mathbf r\tau}$ and $\mu$
should be determined self-consistently from the relations
\begin{eqnarray}
\Delta_{{\mathbf r}{\mathbf \tau}}&=&J\langle c_{{\mathbf r}
\uparrow}c_{{\mathbf r}+
{\mathbf \tau}\downarrow}\rangle _{\Delta_{\mathbf r\tau}},
\label{gapeq} \\
n_e&=&{1\over N}\sum_{{\mathbf r} \sigma} \langle c_{{\mathbf r}
\sigma}^{\dag} c_{{\mathbf r}\sigma}
\rangle _{\Delta_{\mathbf r\tau}},
\label{numbereq}
\end{eqnarray}
where J is the coupling constant
(assumed disorder independent), $N$ is
the system size, and
$\langle  A \rangle _{\Delta_{\mathbf r\tau}}$
means the average of
A in the ground state of $H[\Delta_{\mathbf r\tau}]$.
Without disorder
$\Delta_{{\mathbf r}{\mathbf \tau}}$ is independent of ${\mathbf r}$
and has a particular symmetry with respect to ${\mathbf \tau}$.
In this paper only the on-site
s-wave pairing state $\Delta_{\mathbf r\tau}
=\Delta \delta_{\mathbf \tau 0}$ and the d-wave pairing state
$\Delta_{\mathbf r\tau}=\Delta (\delta_{{\mathbf \tau}\pm \hat x}-
\delta_{{\mathbf \tau}\pm \hat y})$
in two dimensions will be considered.

The Hamiltonian (\ref{ee1}) is bilinear in fermion operators and
can be diagonalized by solving a one-particle problem.
$\Delta_{\mathbf r\tau}$
and $\mu$ should be determined self-consistently from
Eqs. \ref{gapeq} and \ref{numbereq}, which
amount to solving the Bogolubov-de Gennes equations for the
disordered superconductor. For simplicity in calculation,
we perform a particle-hole
transformation for the down-spin electrons,
i.e. $c_{{\mathbf r}\downarrow}
\longleftrightarrow c_{{\mathbf r}\downarrow}^\dagger$,
and re-express (\ref{ee1}) as
\begin{eqnarray}
H^\prime [\Delta_{\mathbf r\tau} ]& =&
-t\sum_{\langle {\mathbf r}{\mathbf r}^\prime \rangle }
(c_{{\mathbf r}\uparrow}^\dagger c_{{\mathbf r}^\prime\uparrow}
- c_{{\mathbf r}\downarrow}^\dagger
c_{{\mathbf r}^\prime\downarrow})
+\sum_{{\mathbf r}{\mathbf \tau}}
(\Delta_{{\mathbf r}{\mathbf \tau}}
c_{{\mathbf r}\uparrow}^\dagger
c_{{\mathbf r}+{\mathbf \tau}\downarrow} +
h.c.)\nonumber\\
&&+\sum_{{\mathbf r}}(\sum_{{\mathbf r}_i}
V_{{\mathbf r}_i,\mathbf r}-\mu )
( c_{{\mathbf r}\uparrow}^\dagger c_{{\mathbf r}\uparrow}
-c_{{\mathbf r}\downarrow}^\dagger
c_{{\mathbf r}\downarrow})+const.
\label{ee2}
\end{eqnarray}
$H^\prime$ has the usual tight binding model form,  but
the hopping constant, the chemical potential, and the
impurity potential have opposite signs for the up spin electrons and
the down spin electrons\cite{half}.
In the remainder of the paper we set the hopping constant t=1.

In Sec. I\/I, the numerical method and essential approximation
is outlined. Results for
the disorder dependence of the zero temperature gap and
superfluid density $\rho_s$ as well as
the quasiparticle density of states $\rho$ are presented.
In Sec. I\/I\/I, a concluding remark is given.

\section{Numerical method and results}

\subsection{Gap parameter and superfluid density}

In the presence of disorder, the energy gap
is space dependent. We determine
$\Delta_{\mathbf r\tau}$ by iteratively solving $H^\prime$
with the self-consistent conditions.
We start from an initial gap function $\Delta_{\mathbf r\tau}$
with a certain pairing symmetry. After diagonalizing $H^\prime$,
we find a new gap function $\Delta_{\mathbf r\tau}$
from the self-consistent equations and then use it as input to
repeat the above
process until the self-consistent conditions are satisfied.
This is a strict self-consistent iterative process.
However, since the gap function at every site needs be
adjusted to satisfy the self-consistent equations,
this is excessively time consuming when an average over a large number
of impurity configurations is required.
We shall however perform this strict
self-consistent iteration procedure
only for studying the properties of a single impurity system and
for checking the accuracy of the approximation used
in many-impurity cases.

As mentioned above, to solve the self-consistent equations,
the Hamiltonian needs be exactly diagonalized.
This can be done, however,
only on small lattices in the presence of disorder. For a
superconductor, a characteristic length scale is the
superconducting correlation length $\xi\sim {\hbar}v_F/\pi\Delta$.
If $\xi$ is larger than the dimension of the system,
the finite size effect is large and the analysis
of the disorder effect may be subtle. To
avoid this situation,  we shall limit our calculations
only to cases where
$\xi$ is much smaller than the dimension of the system.

The existence of a finite superfluid density $\rho_s$
is a defining property of superconductors.
Experimentally $\rho_s$ is determined
from the microwave measurement of the penetration depth.
$\rho_s$ on a finite lattice can be evaluated directly
from the current-current correlation function,
using the eigenfunctions obtained from the iteration procedure
described above, \cite{super}
\begin{equation}
{\rho_s \over 4} =\langle -K_x\rangle
-\Lambda_{xx}(q_x=0,q_y\rightarrow 0,\omega =0)
\end{equation}
where $\langle A\rangle =Tr(Ae^{-\beta H})/Tre^{-\beta H}$,
$\langle K_x\rangle $ is the kinetic
energy along x-direction, and
\begin{eqnarray}
\Lambda_{xx}({\mathbf q},\omega )&=& {i\over V}\int_{-\infty}^tdt'
e^{i\omega (t-t')}\langle
[J_x^p(-{\mathbf q},t),J_x^p({\mathbf q},t')]\rangle
\nonumber\\
&=&{1\over V}\sum_{n_1\not= n_2}{A_{n_1,n_2}(-q)A_{n_2,n_1}(q)
\over \omega +i\delta +E_{n_1}-E_{n_2}}
(\langle c_{n_1}^\dagger c_{n_1}\rangle
- \langle c_{n_2}^\dagger c_{n_2}\rangle  )
\end{eqnarray}
with $c_n = \sum_{{\mathbf r}\sigma }
\phi_n({\mathbf r},\sigma ) c_{{\mathbf r}\sigma }$
the quasiparticle operator and
\begin{equation}
A_{n,m}({\mathbf q})= \sum_{{\mathbf r}\sigma}
e^{i{\mathbf q}\cdot {\mathbf r}}(
\phi_n({{\mathbf r}+x\sigma }) \phi_m({{\mathbf r}\sigma })
-\phi_n({{\mathbf r}\sigma }) \phi_m({{\mathbf r}+x\sigma })).
\end{equation}

Before studying disordered systems, we consider a one-impurity
system. We first consider the change of the gap function
induced by an impurity.
Fig. \ref{fig1} shows the self-consistent energy gap
$\Delta_{\mathbf r \tau}$ for a s-wave superconducting state
with one impurity on a 21$\times$21 lattice.
Since the scattering potential is short ranged, the gap
function changes only in the vicinity of the impurity.
Far away from the impurity, $\Delta_{\mathbf r\tau}$
approaches to the value of the energy gap without disorder.
In a region with a length scale of the range of scattering potential
around the impurity site $\Delta_{\mathbf r}$ is largely reduced
due to the strong suppression of
the probability of an electron hopping
to this region by the impurity scattering.
Beyond this region, there exists a relatively larger
region with a length
scale comparable with the superconducting correlation
length $\xi$ where
a weak oscillation of $\Delta_{\mathbf r\tau}$
in space is observed. This oscillation is due to the
interplay between the impurity scattering and the superconducting
correlations.
Because of the lattice effect, $\Delta_{\mathbf r\tau}$ is
not isotropic in space.
It is relatively strong along the diagonal direction.
Along other directions, it is too weak to be resolved
from the figure. For the d-wave state, similar results have
been found. But the oscillation of $\Delta_{\mathbf r\tau}$
along two axes seems more apparent in this case.

For finite impurity doping systems where many
impurity configurations are used in the disorder average
we shall approximate $\Delta_{{\mathbf r}{\mathbf \tau}}$ in
(\ref{ee1}) by the average of the pairing correlation functions
$\langle c_{{\mathbf r}\downarrow}
c_{{\mathbf r}+{\mathbf \tau}\uparrow}\rangle$ in space,
${\bar\Delta}_{\mathbf \tau}$,
so that only a simplified self-consistent equation,
\begin{equation}
\bar{\Delta}_{\tau}
={1\over N}\sum_{\mathbf r}J\langle c_{{\mathbf r}\downarrow}
c_{{\mathbf r}+{\mathbf \tau}\uparrow}\rangle_{{\bar\Delta}_\tau}
\end{equation}
needs be solved. This approximation is to ignore the fluctuation
of $\Delta_{\mathbf r\tau}$
(but not $\langle c_{{\mathbf r}\downarrow}
c_{{\mathbf r}+{\mathbf \tau}\uparrow}\rangle$) in space.

The errors resulting from the above
approximation can be found by directly
comparing the results obtained with and without the approximation.
We have calculated the errors for
the pairing amplitude for
several arbitrarily chosen configurations of impurities
in both strong and weak scattering potential limits.
For all the cases we have studied, we find that the relative errors
in the average energy gap are small compared with the
combined errors produced by the disorder average or the finite
size effect. For example, for a randomly chosen system of 3 impurities
on a 14$\times$14 lattice, the relative errors in the average
energy gap are 0.1\% (0.6\%) in a weak potential $V_0=2$ and $V_1=0$
and 2\%(2\%) in a strong potential $V_0=20$ and $V_1=0$ for
the s-wave (d-wave) pairing state.
As the impurity concentration increases, the error increases slightly.
The errors for the local pairing correlation functions
$\langle c_{{\mathbf r}\downarrow}
c_{{\mathbf r}+{\mathbf \tau}\uparrow}\rangle$
are larger than that for the average gap, but still small.
Fig. \ref{fig2}, as an example, shows the relative
error pattern for the local correlation function
$\langle c_{{\mathbf r}\downarrow}
c_{{\mathbf r}+{\mathbf \tau}\uparrow}\rangle$
for a s-wave pairing state.
We find that the errors for $\langle c_{{\mathbf r}\downarrow}
c_{{\mathbf r}+{\mathbf \tau}\uparrow}\rangle$
are largest (about 5\%)
at the impurity sites. Clearly $\langle c_{{\mathbf r}\downarrow}
c_{{\mathbf r}+{\mathbf \tau}\uparrow}\rangle$ is
overestimated in the vicinity of impurities and underestimated
far away from the impurities. Nevertheless it is
encouraging that the error made in neglecting off-diagonal disorder
is small\cite{lambert}.

Using the above approximation, we have evaluated
the average gaps $\bar\Delta$
for both s- and d-wave pairing states in different scattering
potentials. For all the cases we have studied,
we find that $\bar\Delta$
decays almost linearly with x for small x.
Fig. \ref{fig3} shows
${\bar\Delta}_\tau$ as a function of the impurity concentration
x for both pairing states in a strong $\delta$-function
scattering potential $V_0=20$ on a 14$\times$14 lattice.
All four curve shown in the figure are nearly parallel to each other
although their values at x=0, i.e. $\bar\Delta (0)$,
are quite different.
This remarkable behavior indicates that
the reduction of $\Delta$ by disorder ($d\Delta / dx$) is
determined only by the scattering potential to
a first approximation and is independent of the
pairing symmetries and the values of the energy gaps for
pure samples.
This nearly universal property of the energy gap could be
of use in future studies of disordered superconductors.
For weak $\delta$-function potentials or finite but short ranged
potentials, similar results for $\bar\Delta$
have been found. But for a weak $\delta$-function potential the slope
of the decay of $\bar\Delta$ with x becomes smaller.

Our calculations for $\bar\Delta$ and $\rho_s$
have been done mostly on lattices with size ranged from
10$\times$10 to 18$\times$18 sites.
In general, $\bar\Delta$ and $\rho_s$ are size dependent.
However, for the quantities which are physically more interesting,
the relative energy gap $\bar\Delta (x)/\Delta (0)$ and the relative
superfluid density $\rho_s (x)/\rho_s (0)$, the finite size effect
is small. A comparison for $\bar\Delta (x)/\Delta (0)$ and
$\rho_s (x)/\rho_s (0)$ for the s- and d-wave states with a
$\delta$-function potential $V_0=8$ on
lattices of 10$\times$10, 14$\times$14, and 18$\times$18 sites
is given in Fig. \ref{fig4}.
Similar results have been found for other impurity potentials.
In general, we find that
the finite size effect for $\rho_s$ is larger than that for
${\bar\Delta}_\tau$ when ${\bar\Delta}_\tau$ is small (i.e.
large $\xi$). Physically this is because that
$\Delta_{\mathbf r\tau}$ is determined only by the local pair
correlation function while $\rho_s$ is determined by the
current-current correlation function at long wavelengths
which in turn is determined by the properties  of the eigenfunction
on the whole lattice.

Fig. \ref{fig5} shows $\bar{\Delta} (x)/\bar{\Delta} (0)$ and
$\rho_s (x)/\rho_s (0)$
as functions of $x$ for the s- and d-wave pairing states in three
potentials.  For a weak $\delta$-function potential, $\bar{\Delta}
(x)/\bar{\Delta} (0)$
and $\rho_s (x)/\rho_s (0)$ decrease very slowly with $x$ and
the difference between the s- and d-wave pairing states is also small.
With increasing $V_0$, the difference between these two pairing
states increases. The disorder has stronger effect on the d-wave
state than the s-wave state; in particularly, $\rho_s$ falls much
faster in the d-wave state than in the s-wave state.
This property may be useful for distinguishing
a d-wave pairing state from
a s-wave pairing state in the unitary scattering
limit from an experimental  point of view.
But if an off-site scattering potential (i.e. $V_1$ term) is present,
this difference will be eventually reduced.
The disorder effect is clearly strongly enhanced
in a finite range potential than
in a $\delta$-function potential.

Fig. \ref{fig6} shows $\bar{\Delta} (V_0,V_1)/\bar{\Delta} (0,0)$
as functions of $V_0$ for the s-
and d-wave pairing states with $x$=0.02. The qualitative behaviors of
$\bar{\Delta} (V_0,V_1)/\bar{\Delta} (0,0)$
are similar for all the cases shown in the figure.
$\bar{\Delta} (V_0,V_1)/\bar{\Delta} (0,0)$
decreases with $V_0$ for small $V_0$, but soon becomes saturated
when $V_0$ surpasses the band width.
 $d\bar{\Delta}/dx$ for the s- and d-wave states
can be very large depending on the value of $V_1$.

\subsection{Density of states}
Now we consider the effect of disorder on the density of states
of quasiparticles. We calculate the density of states using a
recursion method\cite{recursion}. This method
addresses the local density of states of an infinite lattice.
In this method, the density of states $\rho (E)$ is obtained from the
imaginary part of the one-particle Green's function $G(E)$.
\begin{equation}
\rho (E)=\lim_{\epsilon\longrightarrow 0}-{1\over \pi} {\hbox {Im}}
 G(E+i\epsilon).
\end{equation}

Given a starting state $|0\rangle$,
the recursion method is defined by recurrence relations
\begin{equation}
H|0\rangle = a_0 |0\rangle + b_1 |1\rangle \label{rec1}
\end{equation}
and
\begin{equation}
H|n\rangle = b_n |n-1\rangle +a_n |n\rangle + b_{n+1} |n+1\rangle
\qquad (n >0), \label{rec2}
\end{equation}
where $\{ |n\rangle \}$ is a set of normalized bases generated
automatically from equations (\ref{rec1}) and (\ref{rec2}).
{}From the a's and b's generated, $G(E)$ can
be expressed in a continued fraction form
\begin{equation}
G(E)={1\over\displaystyle (E-a_0)-{\displaystyle b_1^2\over
\displaystyle (E-a_1)-{\displaystyle b_2^2\over
\displaystyle (E-a_2)- \cdot\cdot\cdot}}}.
\end{equation}
In real calculation, this continued fraction is truncated
at a certain step and the remainder of the continued fraction
is replaced by a parameter which is determined such that
the error is minimized.
We truncate the continued fraction at a step
when the difference between the result obtained at that step and that
with 5 more steps is smaller than the error demanded.
In using the recursion method, the values of $\Delta_{\mathbf r\tau}$
and $\mu$ obtained previously on finite lattices will be used.
When the impurity concentration is finite, the
approximation $\Delta_{\mathbf r\tau}={\bar\Delta}_\tau$ is
assumed. For one impurity system, the strict
self-consistent solution for
$\Delta_{\mathbf r\tau}$ on a small lattice
around the impurity is used,
while for the rest part of the lattice $\Delta_{\mathbf r\tau}$
are approximated by the average value of $\Delta_{\mathbf r\tau}$
on the edges of the small lattice.

As discussed by Byers et al\cite{byers,choi}
the densities of states in the vicinity of the impurity
is in principle measurable via the spatial variation of the
tunneling conductance around an impurity with a
scanning-tunneling-microscope study of the surface of a
superconductor.
In continuum space, the local density of states
(or the tunneling conductance) around an impurity
has been calculated by Byers et al\cite{byers} for both the s- and
d-wave pairing states and by Choi\cite{choi} for the d-wave pairing
state. They find that the density of states for a
give energy oscillates in space and
depends strongly on the anisotropy of the gap parameter.
When the energy is larger than $\Delta$, the oscillation is largest
in the directions of the gap maxima and smallest in the directions of
the gap minima. In lattice space, however, we find that their
results are partly altered.
Fig. \ref{fig7} shows the impurity induced density of states
as a function of distance from an impurity along two directions for
the s- and d-wave pairing states.
(Here the fully self-consistent isolated
impurity result for the gap function on a
21 $\times$ 21 lattice shown in
Fig. \ref{fig1} is used as
an input.) The density of states oscillates in space with an
energy and direction
dependent wavelength in agreement with the results
of Ref. \cite{byers}.
In the s-wave case, the oscillation along the
diagonal direction is much larger than that along the x-axis direction,
in contrast to the isotropic s-wave pairing state in continuum space.
Since the energy gaps are the same on the Fermi surface for a
s-wave states, this
difference is purely a lattice effect.
For the d-wave pairing state, the oscillations
along two directions are not so different as shown in Refs.
\cite{byers,choi}. The impurity induced density of states decays
slightly faster along the x-axis direction than along the
diagonal direction.

To compute the density of states with finite doping of impurities,
we have used the results of the average gap $\bar\Delta$ obtained
previously. The fluctuation of $\Delta_{\mathbf r\tau}$
in space is ignored in this calculation.
Fig. \ref{fig8} shows the density of states for the s- and
d-wave pairing states with different
potentials and x. The main results
are summarized as follows:

(a) For s-wave pairing with weak potential or strong potential with
very small x, $\rho$ has no qualitative change with respect to
the case without disorder. In particularly,
the energy gap still exists and
is hardly changed by disorder in agreement
with Anderson theorem\cite{pwa}.

(b) For the d-wave pairing state with weak scattering potential,
the change of $\rho$
with respect to the case without disorder is small. But $\rho$ at
the Fermi energy $E_F$ becomes finite,
in consistent with the non-magnetic
impurity scattering theory in the Born scattering limit\cite{GK}.

(c) For the d-wave pairing state with very strong
onsite potential, $\rho$ shows a peak around $E_F$.
This results agrees very well with the self-consistent t-matrix
theory for the d-wave superconductor in the
unitary scattering limit\cite{PEAK}. On the other hand it also
shows that the approximations made
in the self-consistent t-matrix theory,
such as ignoring the vertex corrections and
the energy dependence of the
self-energy, are valid for the d-wave state.

(d) For the d-wave pairing state with a strong on-site potential or
both pairing states with a finite range potential, $\rho$ at $E_F$
grows quickly with increasing x and becomes comparable with the
average density of states at some critical x.
For the s-wave pairing state with a strong $\delta$-function
potential, a finite gap remains when
x is smaller than a critical value x$_c$ within numerical errors.
When x$>$x$_c$, the gap vanishes (but $\rho_s$ and ${\bar\Delta}$
are non-zero),
$\rho$ at $E_f$ is small and increases slowly
with increasing x.

(e) All singularities of $\rho$
are suppressed by disorder average in these calculations.
We have not found any evidence for the singular behavior
predicted recently by Nersesyan et al\cite{ntw}
within numerical error.

\section{Conclusion}

We have discussed a straightforward numerical technique which allows
detailed effects of various impurity potentials on
superconductors to be investigated with a BCS mean field framework.
In particular we evaluated the gap parameter,
the superfluid density, and the density of states for the
s-wave and d-wave superconducting states with non-magnetic impurities,
as functions of impurity concentrations and scattering potentials.
For one impurity systems, the local
density of states induced by impurity oscillates in space,
in agreement with known analytic results.
For s-wave pairing, the energy gap and the
density of states are hardly affected by weak disorder
(i.e. either weak scatterers or dilute strong scatterers),
consistent with Anderson theorem.
In dilute impurity limit, our results agree well
with the self-consistent t-matrix theory in both Born and unitary
scattering limits, and in both s- and d- pairing states.
For the d-wave pairing state,
the density of states at $E_F$ becomes finite even for
weak scattering potential, consistent with the Born scattering
theory of the d-wave superconductor.
For the d-wave pairing state
with strong on-site potential, the density
of states is in good agreement with the self-consistent t-matrix
theory for the d-wave superconductor in the unitary limit.
For strong scatterers, the energy gap of the s-wave
state disappears beyond a critical doping level
which is sensitive to the
range of the impurity potential.
A finite range potential is shown to have a stronger effect than a
short range potential in either pairing state.
For Zn doped YBaCuO, experiments find that $T_c$ varies almost
linearly with $x$ and drops about 25\% for 2\% Zn doping\cite{TC}.
If we assume the change of $T_c$ is equivalent to the change of
$\bar{\Delta}$ at zero temperature, we find that
$d\bar{\Delta}/dx$ in a $\delta$-function
potential is too small to fit quantitatively with experiments
even in unitary scattering limit. However for a finite
range potential,
no such difficulty exists.
\section{Acknowledgement}
We wish to thank J. Loram, J. Cooper, and
P. Hirschfeld for helpful communications.

\newpage
\begin{figure}
\caption{Self-consistent gap function
for the s-wave superconducting state on a 21$\times$21
lattice. Only half of the lattice is shown. The
impurity is located at the center of the lattice.
\label{fig1}}
\end{figure}

\begin{figure}
\caption{Relative errors of the local pairing correlation function
$({\langle c_{{\mathbf r}\downarrow}c_{{\mathbf r}\uparrow}
\rangle}_{{\bar\Delta}}-
{\langle c_{{\mathbf r}\downarrow}c_{{\mathbf r}\uparrow}
\rangle}_{{\Delta}_{\mathbf r\tau}})/
\overline {\langle c_{{\mathbf r}\downarrow}c_{{\mathbf r}\uparrow}
\rangle}$
(where ${\langle c_{{\mathbf r}\downarrow}c_{{\mathbf r}\uparrow}
\rangle}_{\bar\Delta}$
and ${\langle c_{{\mathbf r}\downarrow}c_{{\mathbf r}\uparrow}
\rangle}_{{{\Delta}_{\mathbf r\tau}}}$ are the results obtained with
and without approximation, and
$\overline{\langle c_{{\mathbf r}\downarrow}c_{{\mathbf r}\uparrow}
\rangle}$ is the average of ${\langle c_{{\mathbf r}\downarrow}
c_{{\mathbf r}\uparrow}
\rangle}_{{\Delta}_{\mathbf r\tau}} $ in space)
for the s-wave pairing state on a 14$\times$14
lattice with 3 impurities and a potential
$V_0=20$ and $V_1=0$. The impurities are located on the sites where
the three highest peak emerge.
\label{fig2}}
\end{figure}

\begin{figure}
\caption{Energy gap ${\bar\Delta}$ vs x for both pairing states with
the coupling constant J=1.5 and J=2.3 and the scattering potential
$V_0=20$ and $V_1=0$ on a 14$\times$14 lattice.
25 impurity configurations are used in disorder average.
\label{fig3}}
\end{figure}

\begin{figure}
\caption{Normalized energy gap $\bar\Delta (x)/\bar\Delta (0)$ and
normalized
superfluid density $\rho_s (x)/\rho_s (0)$ as functions of impurity
concentration x on the square lattices of 10$\times$10, 14$\times$14,
and 18$\times$18 sites with $V_0=8$ and $V_1$=0. $\bar\Delta (0)$
is about 0.5 (0.3) for
the s- (d-) wave pairing state, which is about 1/20 band width.
100 impurity configurations for 10$\times$10 lattice,
and more than 25 impurity configurations
for 14$\times$14 and 18$\times$18 lattices
are used in disorder average.
\label{fig4}}
\end{figure}

\begin{figure}
\caption{Normalized energy gap
$\bar\Delta (x)/\bar\Delta (0)$ and
normalized superfluid density $\rho_s (x)/\rho_s (0)$
as functions of
x for both s-wave and d-wave superconducting states
with three different potentials on 10$\times$10 lattices.
$\bar\Delta (0)$ are the same as for Fig. \ref{fig4}.
100 configurations
of impurities are used in disorder average.
\label{fig5}}
\end{figure}

\begin{figure}
\caption{Normalized energy gap $\bar\Delta (V_0,V_1)/\bar\Delta (0,0)$
as functions of
$V_0$ for the s- and d-wave pairing states with x=0.02. The values of
$\bar\Delta (0,0)$ are the same as $\bar\Delta (0)$
used in Fig. \ref{fig4}.
100 configurations of impurities are used in disorder average.
\label{fig6}}
\end{figure}

\begin{figure}
\caption{The local density of states induced by an impurity
as a function of distance r from an impurity
along both the x-axis direction ($0^\circ$)
and the diagonal direction ($45^\circ$)
at energy $\omega =1.5\Delta_{max}$. $V_0=2$ and $V_1=0$.
\label{fig7}}
\end{figure}

\begin{figure}
\caption{Density of states as functions of energy E for s- and d-wave
superconducting states with different impurity concentrations x.
The parameters $\bar\Delta$ obtained in Fig. \ref{fig5} are
used in calculations here.
More than 1000 configurations of
impurities are used in disorder average.
\label{fig8}}
\end{figure}


\begin{references}
\bibitem{phaseshifts} It is worth noting some features of this 2D
scattering problem.
Using the impenetrable disc model
with radius $a$ we find that $k_f a=0.542$
satisfies the sum rule with near unitary scattering in the
$l=0$ channel; the corresponding phase shifts are
$\delta_0 = -0.75 {(\pi / 2)}$, $\delta_1 =-0.12{(\pi / 2)}$,
$\delta_2 = -0.005{(\pi / 2)}$ etc.
The next values of $k_f a$ compatible with the sum rule
are $1.52$ and $2.787$.
The latter value leads to scattering phase shifts
$\delta_0 = -0.25 {(\pi / 2)}$,
$\delta_1 = 0.64{\pi / 2}$, $\delta_2 = -0.69{(\pi / 2)}$,
$\delta_3 =-0.26{(\pi / 2)}$ and $\delta_4 = -0.06{(\pi / 2)}$;
they are appreciable in all channels up to $l=2$.
This set of phase shifts leads to the best agreement with the
$YBCO$ data.
\bibitem{half} The results presented here are for a half-filled band.
For the simple tight-binding model used here, this
has a van Hove singularity at the Fermi surface in the normal state.
This fact plays no
role in our $T=0$ calculations however. To see this note that (a)
in the absence of disorder, the superconducting gap pushes the
van Hove singularity far from the Fermi surface, (b)
in the presence of disorder,
all singularities in the density of states are rapidly
suppressed after the
disorder average, even in the normal state.
The chemical potential $\mu$
must be determined self-consistently along with $\Delta$ even
at half filling, because particle-hole
symmetry is destroyed by disorder.
\end{references}
\end{document}